# Ultralow lattice thermal conductivity and electronic properties of monolayer 1T phase semimetal SiTe$_2$ and SnTe$_2$

Yi Wang[1], Zhibin Gao[1] and Jun Zhou[*]

Center for Phononics and Thermal Energy Science, China-EU Joint Center for Nanophononics, Shanghai Key Laboratory of Special Artificial Microstructure Materials and Technology, School of Physics Sciences and Engineering, Tongji University, Shanghai 200092 China

**Abstract**

2H phase (trigonal prismatic D$_{3h}$) of layered two-dimensional (2D) transition metal dichalcogenides (TMDs) have attracted a lot of interests due to the superior electronic and optoelectronic properties. However, flexible electronic devices and thermoelectric performances based on 2H phase have been potentially limited by the strain sensitive electronic band gap and high lattice thermal conductivity ($\kappa_L$). Here, we predict and calculate two 1T (octahedral O$_h$) phase monolayer telluride materials SnTe$_2$ and SiTe$_2$ with soft mechanics, ultralow $\kappa_L$ and electronic properties. The calculated in-plane Young's modulus of monolayer SnTe$_2$ is softer than most of 1T-MX$_2$ compounds. Furthermore, monolayer SiTe$_2$ and SnTe$_2$ also have relatively flexible electronic properties under large biaxial strain, indicating potential flexible electrode materials. Meanwhile, monolayer SiTe$_2$ and SnTe$_2$ both exhibit ultralow $\kappa_L$ (2.27 W/mK of SiTe$_2$ and 1.62 W/mK of SnTe$_2$) at room temperature. Considering both acoustic and polar optical phonon scattering of the electronic relaxation time, the figure of merit (*ZT*) can achieve 0.46 at 600 K and 0.71 at 900 K for monolayer SiTe$_2$ and SnTe$_2$ respectively.

**Keywords**: Monolayer 1T phase SiTe$_2$ and SnTe$_2$; lattice thermal conductivity; electronic properties

[1] These authors contributed equally to this work.
*Corresponding author. E-mail: zhoujunzhou@tongji.edu.cn.

# 1. Introduction

Recently, transitional metal dichalcogenides (TMDs) have attracted a lot of interests due to the wide range of applications in nanoelectronics [1], optoelectronics[2][3], photovoltaics and photodetection [4]. Generally, 2D TMDs have two stable phases that are honeycomb (H) and central honeycomb (T) configurations. Besides, another phase 1T' with lower symmetry, which is a distorted version of 1T structure [5]. At normal condition, 2H phase is more stable than 1T and 1T' and shows superior properties. For example, the well-known monolayer 2D $MoS_2$ of 2H structure is a direct band gap semiconductor, but 1T counterpart is a metal instead. However, Metallic 1T $MoS_2$ has been confirmed as a superior supercapacitor electrode material [6]. Furthermore, a series of 1T-type 2D TMDs have also been reported to have in-plane negative Poisson's ratio due to the strong coupling between different electronic orbits [7]. Even though 1T phase is less stable than 2H, 1T TMDs have been experimentally synthesized [8][9] and have many exotic novel properties [1][7][10].

In 2012, around 44 different combinations of $MX_2$ compounds, including 1T and 2H phases monolayer transition metal oxides and dichalcogenides have been predicted by Ataca *et al* [11]. In all these $MX_2$, element M comes from Group IIIB to VIIIB (like Sc, Ti, V, Cr, Mn, Fe, Co, Ni) and X from group VIA (including S, Se and Te). Hence, it is interesting to explore other stable 1T $MX_2$ with novel properties. Here, we consider the group IVA element M and group VIA element X forming novel 1T-type $MX_2$ monolayer 2D materials. Within this combination, Yu *et al* successfully exfoliated monolayer 1T phase $SnSe_2$ via mechanical exfoliation [8], which has a high *ZT* of 0.94 at 600 K [10]. What's more, 1T-$SnS_2$ is also utilized in high-performance top-gated field-effect transistors (FETs) and related logic gates [9]. Recently, atomically thin 2D tellurium has been reported to have ultralow $\kappa_L$ of 2.16 W/mK [12] and has been synthesized with superior electronic mobility [13][14]. In this sense, monolayer telluride materials would be potential thermoelectric materials with high *ZT*.

A silicon based layered nanostructures of silicon telluride $Si_2Te_3$ have been synthesized successfully [15]. After that, bulk material $SiTe_2$ has also been found via quantum-chemical techniques [16] and the electronic properties of monolayer $SiTe_2$ have been studied [17]. In this paper, we would like explore to some new 1T phase telluride materials with many excellent mechanical and thermoelectric properties. Good thermoelectric materials can directly convert waste heat into electrical energy with high efficiency, which is crucial for the sustainable development under energy crisis and global warming. An optimal thermoelectric material needs to balance a variety of conflicting parameters, which is expressed as $ZT = \sigma S^2/(\kappa_e + \kappa_L)$. To maximize *ZT* of material, a larger Seebeck coefficient (*S*), high electronic conductivity (σ), low electronic thermal conductivity ($\kappa_e$) and $\kappa_L$ are acquired simultaneously [18][19]. In this work, we explore the electronic, mechanical and thermoelectric properties of monolayer 1T phase $SiTe_2$ and $SnTe_2$. We found that at room temperature, monolayer $SiTe_2$ and $SnTe_2$ have ultralow $\kappa_L$ of 2.27 W/mK and 1.62 W/mK respectively. The value of monolayer $SnTe_2$ is smaller than atomically thin 2D tellurium (2.16 W/mK) [12]. They also have relatively stable electronic band structures and density of states (DOS) under biaxial strain, indicating potential flexible electrode materials. Furthermore, we also obtained the maximum *ZT* of 0.46 for monolayer SiTe$_2$ at 600 K and of 0.71 for monolayer SnTe$_2$ at 900 K by considering the scattering between acoustic phonons and polar optical phonons.

## 2. Computational details

All of our calculations of monolayer 1T-XTe$_2$ (X=Si, Sn) are performed using the density functional theory (DFT), as implemented in the Vienna Ab initio Simulation Package (VASP) [20][21] within the projector augmented wave (PAW) [22][23] method. Perdew-Burke-Ernzerhof (PBE) [24] is used as pseudopotential with the generalized gradient approximations (GGA) for exchange correlation functional. 500 eV kinetic energy cut-off for the plane-wave basis set are adopted in computations. The structures convergence criterion is selected as $10^{-10}$ eV for ions and electrons. They are optimized that until a maximum Hellmann-Feynman force on each atom is less than 0.001 eV/Å. In addition, the Monkhorst-Pack k-point grid of 15×15×1 is used with the Γ symmetry to sample the Brillouin zone. A 30 Å vacuum space is applied to the out-of-direction in order to avoid the interactions between adjacent layers. The phonon dispersion curves of monolayer 1T-XTe$_2$ (X= Si, Sn) structures are all calculated using the density functional perturbation theory (DFPT), as implemented in the PHONOPY package [25] interfaced with VASP [20][21]. The lattice thermal conductivity is obtained utilizing the Boltzmann transport equation, as implemented in the shengBTE codes [26][27]. We calculate the second and third anharmonic force constants via using a 6×6×1 supercell, considering the third nearest neighbors. The electronic transport properties are calculated via using Boltzmann transport theory and relaxation time approximation (RTA), as implemented in the BoltzTraP codes [28] based on the electronic band structures.

## 3. Results and discussion

### 3.1 Structure optimization and stability

We calculated two 1T phase monolayer telluride XTe$_2$ (X = Si, Sn) which belong to the space group p$\bar{3}$m1 (NO.164), as shown in Fig 1. From the top view, the optimized structure can be regarded as a superposition of central atom and honeycomb structure of TMDs. It can also be viewed that the positive single-layer hexagonal lattice of X atom is sandwiched between two negatively charged Te atoms in the side view [11]. As we know, the lattice constant of bulk SiTe$_2$ in the experiment is a = b =

7.43 Å, c = 13.48 Å within the trigonal space group p$\bar{3}$c1 (NO. 163) [16]. Monolayer SiTe$_2$ have been studied recently [17]. But 3D and 2D material of SnTe$_2$ have not been caught much attention yet. The optimized structure in Table 1 show that the lattice constant of primitive cell of monolayer SiTe$_2$ and SnTe$_2$ are a = b = 3.78 Å and a = b = 4.10 Å respectively. Meanwhile, the bond length between X atom and Te atom of XTe$_2$ (X = Si, Sn) are 2.73 Å and 2.99 Å. The bond angle of them are 87.52° and 86.82°, respectively.

The intrinsic thickness of structures is exhibited outside the parentheses in Table 1. Values in the parentheses are the effective thickness, considering the van der Waal (vdW) radius of elements [33][34]. Here, the intrinsic thickness of both phases are 3.29 Å and 3.63 Å by structure optimization. The last column $E_c$ is the cohesive energy in the unit cell. The values are calculated using the formula $E_c = [E(X) + 2E(Te) - E(XTe_2)]$ [29][30]. In which $E$(XTe$_2$), $E$(X) and $E$(Te) are the total energy of XTe$_2$ and the total energies of free X and Te atoms respectively. The cohesive energy of monolayer SiTe$_2$ is 15.24 eV and monolayer SnTe$_2$ is 13.83 eV. These values are comparable to the other stable 1T phase telluride materials such as VTe$_2$, MnTe$_2$, NiTe$_2$ and NbTe$_2$ with 14.24 eV, 12.27 eV, 13.19 eV and 16.38 eV respectively [11]. These high cohesive energies reflect good thermal stability of our monolayer 1T XTe$_2$.

The Fig. 2 show the phonon spectra along the high symmetry points Γ-M-K-Γ in the Brillouin Zone (BZ) and DOS of monolayer XTe$_2$. There are nine phonon modes in total, including three acoustic phonon modes (colour lines) and six optical phonon modes (blue lines) due to three atoms in the primitive cell. They are all free of imaginary frequencies in the phonon spectra in Fig. 2(a) and Fig. 2(c), indicating dynamical stability of our XTe$_2$.

3.2 Soft mechanical properties

We calculated the mechanical properties of monolayer SiTe$_2$ and SnTe$_2$, such as the elastic modulus tensor $C_{ij}$, the Young's modulus (in-plane stiffness) and Poisson's ratios. We use the Hooke's law plane-stress condition to calculate the elastic constants and moduli of our 2D materials [31][32]. The formula of Young's modulus and Poisson's ratio can be expressed as: [32][33]

$$E_x = \frac{C_{11}C_{22} - C_{12}C_{21}}{C_{22}}, E_y = \frac{C_{11}C_{22} - C_{12}C_{21}}{C_{11}} \quad (1)$$

$$v_{xy} = \frac{C_{21}}{C_{22}}, v_{yx} = \frac{C_{12}}{C_{11}} \quad (2)$$

The results are shown in Table 2. As the 3D systems can be directly calculated by first-principles theory, but the 2D has to be utilized by the vdW effective thickness of structures in Table 1. They are 7.77 Å and 8.11 Å for monolayer SiTe$_2$ and SnTe$_2$ respectively. In addition, we show the in-plane Young's modulus $E_x$ (or $E_y$) and Poisson's ratio $v_{xy}$ (or $v_{yx}$) in the *x* (or *y*) direction, comparing with other 1T-type telluride materials. In terms of Young's modulus, monolayer TcTe$_2$ is the softest in all 42 1T-MX$_2$ compounds by Yu *et al*. work. The value is 34 N/m [7], but monolayer SiTe$_2$ and SnTe$_2$ are lower than monolayer TcTe$_2$ with 21.16 N/m and 16.80 N/m respectively. To the best of our knowledge, they are the softest 1T phase stable monolayer materials. These soft mechanical properties indicate that our 2D XTe$_2$ materials may have potential in flexible mechanics and electronics. Moreover, we find that monolayer SiTe$_2$, SnTe$_2$ have the positive in-plane Poisson's ratio, the same as monolayer TcTe$_2$ [7].

3.3 Lattice thermal conductivity

As we know, a finite lattice thermal conductivity ($\kappa_L$) is an outcome of the phonon-phonon scattering [35][36]. Our calculated $\kappa_L$ of monolayer SiTe$_2$ and SnTe$_2$ are 2.27 W/mK and 1.62 W/mK at room temperature, which are quite smaller than monolayer SnS$_2$ (6.41 W/mK), SnSe$_2$ (3.82 W/mK) [1] and SnSe$_2$ (3.27 W/mK) [10].

Compared to the monolayer SnSe$_2$, we can illustrate the differences in the following several aspects. Firstly, the studies of Shafique *et al*. show that the maximum frequency of vibration is 249.36 cm$^{-1}$ for monolayer SnSe$_2$ [1]. Meanwhile, there are 305.50 cm$^{-1}$ in monolayer SiTe$_2$ and 162.4 cm$^{-1}$ in monolayer SnTe$_2$. Moreover, the maximum frequency of acoustic phonon modes is about 114.99 cm$^{-1}$ of monolayer SnSe$_2$ [1][10]. Monolayer SiTe$_2$ and SnTe$_2$ are about 99.40 cm$^{-1}$ and 79.53 cm$^{-1}$ in terms of the maximum frequency of acoustic phonon modes. The lower frequency of acoustic phonon modes in monolayer SiTe$_2$ and SnTe$_2$ illustrate that the group velocity of acoustic of them are lower than monolayer SnSe$_2$, indicating ultralow $\kappa_L$ [10]. Secondly, the phonon group velocity of each mode is given by $v_k = \frac{\partial \omega_k(q)}{\partial q}$, where ω, *k*, *q* represent the vibrational frequency, the vibrational mode index and the wave vector respectively [1]. Therefore, we can know that the slope of acoustic phonon modes represents the magnitude of the group velocity approximatively. The acoustic phonon modes of monolayer SiTe$_2$ and SnTe$_2$ are more flat than monolayer SnSe$_2$ from their phonon spectrum and the slope of them are smaller than monolayer SnSe$_2$ [10]. Lastly, the relative atomic mass of the element Te is heavier than that of Se, which will enhance the phonon scattering, and further decrease $\kappa_L$. Similarly, we also know that 1T' phase monolayer MoTe$_2$ and WTe$_2$ have lower $\kappa_L$ with 1.6 W/mK and 1.06 W/mK at room temperature separately [37]. The important reason is associated with heavy atomic mass element of Te. It is also demonstrated that monolayer Te have unusually low $\kappa_L$ because of the weak phonon vibrations and interatomic bonding [12]. As a result, those reasons explain that $\kappa_L$ of monolayer SiTe$_2$ and SnTe$_2$ are lower than monolayer SnSe$_2$.

In Fig. 3(a), temperature dependence of $\kappa_L$ for monolayer SiTe$_2$ and SnTe$_2$ are shown from 300 K to 900 K. The dashed lines show the results of fitting curves of 1/T. It is a good agreement with our calculations when Umklapp process of phonon scattering is dominant the thermal resistivity at a higher temperature. These two structures have ultralow $\kappa_L$ at 300K, 2.27 W/mK for monolayer SiTe$_2$ and 1.62 W/mK for SnTe$_2$. Compared to 2H phase TMDs and other 2D materials, they are extremely low $\kappa_L$ [19][38][39]. Furthermore, Fig. 3(b) show that group velocity of monolayer SnTe$_2$ is smaller than monolayer SiTe$_2$ since the mass of Sn atom is heavier than Si. Moreover, we know that the value of phonon lifetime is determined by the channels of three phonon scattering, which can be calculated by the phase space $P_3$ theoretically [12][27], as shown in Fig. 3(c) and 3(d) respectively. $P_3$ is inversely proportional to phonon lifetime and also related with strength of anharmonicity. The results of Fig. 3(d) show monolayer SnTe$_2$ has strong anharmonicity and it can lead to lower $\kappa_L$. In general, we think that there are in the competition between group velocity and phonon lifetime. $\kappa_L$ of monolayer SnTe$_2$ is lower than SiTe$_2$ because of the domination of phonon lifetime.

3.4 Electronic structures

Electronic band calculations of monolayer SiTe$_2$ and SnTe$_2$ are presented in Fig. 4(a) and 4(c) respectively. The total DOS are shown in Fig. 4(b) and 4(d). As we can know, they are semimetal materials due to the overlap between the conduction band minimum (CBM) and the valence band maximum (VBM) [40]. According to section 3.2 and 3.3, we find monolayer SiTe$_2$ and SnTe$_2$ are much softer than most of other 1T-type monolayer materials. They also have ultralow $\kappa_L$. Thus, we speculate that they may have flexible electronic properties. We consider the changes of band structures and DOS by biaxial strain ε [41]. In Fig. 4, bule, red and black lines represent ε= 5%, ε= -5% and no strain respectively. Around the Fermi level, the changes of electronic band structures and the values of DOS are small. For example, energy changes of CBM of monolayer SiTe$_2$ and SnTe$_2$ vary from ε=0 to ε= 5%. The reduced energies are about 0.35 eV and 0.14 eV respectively. From ε=0 to ε= -5%, the increased energies are about 0.04 eV and 0.13 eV. Importantly, they are still semimetal materials under biaxial strain. Therefore, they have relatively flexible electronic properties. Combined with their soft mechanical properties in section 3.2, it is indicated that monolayer SiTe$_2$ and SnTe$_2$ may be potential flexible electrode materials.

Considering relatively flexible electronic properties, we then focus on exploring their electronic transport properties without strain. Around the Fermi level, the CBM have much higher DOS than the VBM. As a result, we can roughly estimate that moving the Fermi level with a certain carrier concentration could acquire a higher Seebeck coefficient, because it is proportional to the DOS effective mass [42][43].

3.5 Electronic transport properties

On the basis of the electronic structures, the Boltzmann transport theory and the relaxation time approximation (RTA) can be used to calculate the electronic transport properties for monolayer $SiTe_2$ and $SnTe_2$. The electronic conductivity σ and electronic thermal conductivity $\kappa_e$ are all dependent of the relaxation time τ. We focus on considering charge scattering mechanism for monolayer $SiTe_2$ and $SnTe_2$, which plays a critical role in electronic transport properties. Usually, it is reasonable that the carries scattering of acoustic phonon, polar optical phonon and impurities should be considered [54]. Here, we mainly consider acoustic and polar optical phonon scattering.

We can estimate acoustic phonon scattering of the relaxation time τ based on deformation potential theory via the effective mass approximation. The formula of the carrier mobility can be expressed as $\mu_{ac} = \frac{e\hbar^3 C_{2D}}{k_B T m^* m_d E_l^2}$ [1][44]-[51]. Meanwhile, the carrier mobility can also be defined as $\mu_{ac} = \frac{e\tau_{ac}}{m^*}$. Therefore, we can obtain the formula of the relaxation time $\tau_{ac}$,

$$\tau_{ac} = \frac{\hbar^3 C_{2D}}{k_B T m_d E_l^2} \quad (3)$$

According to the formula (3), the relaxation time $\tau_{ac}$ can be calculated approximately, where $e$, $C_{2D}$, $k_B$, T, $m_d$ and $E_l$ are the electron charge, the Planck constant, the elastic constant, the Boltzmann constant, temperature, the average effective mass and deformation potential constant. Here, the average effective mass can be divided into two parts. It should consider the degeneracy of CBM and VBM. The formula is expressed as $m_d = \sqrt{m^*_{\Gamma-M} m^*_{M-K}}$ [1]. The deformation potential constant ($E_l$) is calculated by $E_l = \frac{dE_{edge}}{d\delta}$, where $E_l$ represents the energy of the band edges (VBM for the holes and CBM for the electrons) by uniaxial strain ε [52]. All parameters of monolayer $SiTe_2$ and $SnTe_2$ are listed at 300 K in Table 3.

Furthermore, we consider polar optical phonon scattering to estimate the relaxation time $\tau_{op}$. The equation of mobility for 2D material is referred to $\mu_{op} = \frac{4\pi\varepsilon_0 \varepsilon_p \hbar^2}{e\omega_{LO} m^{*2} Z_0}[e^{\hbar\omega_{LO}/k_B T} - 1]$ [53]-[55] where $\frac{1}{\varepsilon_p} = \frac{1}{\varepsilon_\infty} - \frac{1}{\varepsilon_s}$, $\varepsilon_0$ is the vacuum permittivity, $\varepsilon_\infty$ and $\varepsilon_s$ are the high frequency and static dielectric constant respectively. Meanwhile, $m^*$ is the effective mass of electron. We can obtain this value according to the formula $m^* = \hbar^2(\frac{\partial^2 E(k)}{\partial k^2})^{-1}$, which can be obtained from the curvature of the band edge. The curvature $\frac{\partial^2 E(k)}{\partial k^2}$ is calculated with squares fit to quadratic function [1][52]. Here, the values of monolayer SiTe$_2$ and SnTe$_2$ are 0.068 $m_e$ and 0.19 $m_e$ respectively. $\hbar\omega_{LO}$ is polar optical phonon energy and Z$_0$ is the crystal thickness. In our calculations, we use the intrinsic thickness of structures here [54]. In the limitation of high temperature, we can obtain the formula of the relaxation time $\tau_{op}$,

$$\tau_{op} = \frac{4\pi\varepsilon_0 \hbar^3}{e^2 k_B T m^* Z_0}(\frac{1}{\varepsilon_\infty} - \frac{1}{\varepsilon_s})^{-1} \quad (4)$$

Here, the $\varepsilon_\infty$ and $\varepsilon_s$ for monolayer SiTe$_2$ and SnTe$_2$ are 0.95, 11.45 and 0.97, 8.54 respectively. The results of $\tau_{op}$ for them utilizing the formula (4) are shown in Table 3. Thus, we can estimate the total relaxation time of electron according to the Matthiessen rule, $\frac{1}{\tau_{tot}} = \frac{1}{\tau_{ac}} + \frac{1}{\tau_{op}}$. In Table 3, we obtain $\tau_{tot}$ of SiTe$_2$ and SnTe$_2$ with 1.84×10$^{-14}$ s and 1.56×10$^{-14}$ s at 300 K. From the results, the polar optical phonon scattering plays an important role for carrier relaxation time τ which can overestimate the value of $\tau_{tot}$ without considering the contributions of LO phonon. The *ZT* of material may be overestimated eventually. Based on the estimated values of $\tau_{tot}$, we next calculate all the electronic transport coefficients at different temperatures using the Boltzmann transport equation [28]. We also simulate changes of the carrier concentration utilizing the rigid band approximation (RBA) [10][56][57].

Fig. 5 and Fig. 6 show that all the electronic transport parameters vary with the same range of effective carrier concentration at 300 K, 600 K and 900 K when moving the Fermi level for monolayer SiTe$_2$ and SnTe$_2$ respectively. Importantly, we need to calculate effective carrier concentration of 2D material considering the thickness of unit cell because the default is the value of bulk materials in BoltzTrap codes [58][59]. For the two monolayer materials in Fig. 5(a) and Fig. 6(a), the Seebeck coefficients *S* appear the extreme values, that is about 145.25 μV/K at the effective carrier concentration of 4.53×10$^{14}$ cm$^{-2}$ for SiTe$_2$ at 300 K and 145.30 μV/K at 3.84×10$^{14}$ cm$^{-2}$ for SnTe$_2$ at 900 K respectively. Both are lower than monolayer SnSe$_2$ of 400 μV/K [10]. They also have large σ in Fig. 5(b) and Fig. 6(b) at three temperatures because of the conductivity of the semimetal material. In Fig. 5(c) and Fig. 6(c), the power factor (*PF*) is large at 300 K, because it also has the larger S and σ comparing to 600 K and 900 K. But we know the constraints between S and σ. Finally, we present results of *ZT* at the different temperatures based on above electronic and phononic transport coefficients in Fig. 5(d) and 6(d). It can be found that the maximum value of SnTe$_2$ is reached by 0.71 at 900 K with the effective carrier concentration $3.91\times10^{14}$ cm$^{-2}$, and 0.46 at 600 K with $4.03\times10^{14}$ cm$^{-2}$ for SiTe$_2$.

## 4.Conclusion

In conclusion, we have predicted and calculated 1T phase monolayer semimetal materials of SnTe$_2$ and SiTe$_2$, and explored their mechanical, the phononic and electronic transport properties based on first principles calculation combining with the Boltzmann transport theory. Comparing to other reported 42 1T-MX$_2$ compounds in Ref. [7], monolayer SnTe$_2$ is softer. Monolayer SiTe$_2$ and SnTe$_2$ have relatively flexible electronic properties under biaxial strain ε= ±5%. Combining soft mechanics with flexible electronic properties, these two structures may be potential flexible electrode materials. Furthermore, $\kappa_L$ of monolayer SiTe$_2$ and SnTe$_2$ are lower than monolayer SnSe$_2$ [1][10] that are the lowest in the reported 1T-type structures. We attribute the reasons of ultralow $\kappa_L$ to three aspects: the lower maximum frequency of acoustic phonon modes of monolayer SiTe$_2$ and SnTe$_2$, the lower slope of three acoustic phonon modes with the lower group velocity of monolayer SiTe$_2$ and SnTe$_2$ and the heavier atomic mass of the element Te. Meanwhile, we have obtained the ultralow $\kappa_L$ of monolayer SiTe$_2$ and SnTe$_2$ of 2.27 W/mK and 1.62 W/mK respectively. We also have calculated the electronic transport properties without strain and consider both the acoustic and polar optical phonon scattering to estimate the electronic relaxation time. The results show that *ZT* of monolayer SiTe$_2$ and SnTe$_2$ can achieve 0.46 at 600 K and 0.71 at 900 K.

**Acknowledgements**

We thank Prof. Jie Ren for very helpful discussions, and his critical reading and correcting of the manuscript. This work is supported by the National Natural Science Foundation of China (No. 11775159), the Natural Science Foundation of Shanghai (No. 18ZR1442800), the National Youth 1000 Talents Program in China, and the Opening Project of Shanghai Key Laboratory of Special Artificial Microstructure Materials and Technology.

# References

[1] A. Shafique, B. Samad and Y. H. Shin, *Phys. Chem. Chem. Phys* 19 (2017) 20677-20683.

[2] J.A. Wilson and A. D. Yoffe, *Adv in Phys* 18 (1969) 193-335.

[3] Q. H. Wang, K. Kalantar-Zadeh, A. Kis, J. N. Coleman and M. S. Strano, *Nat Nanotech* 7 (2012) 699-712.

[4] D. B. Velusamy, R. H. Kim, S. Cha, J. Huh, R. Khazaeinezhad, S. H. Kassani, G. Song, S. M. Cho, S. H. Cho, I. Hwang, J. Lee, K. Oh, H. Choi and C. Park, *Nat Commun* 6 (2015) 8063.

[5] K. A. Duerloo, Y. Li and E. J. Reed, *Nat Commun* 5 (2014) 4214.

[6] M. Acerce, D. Voiry and M. Chhowalla, *Nat Nanotech* 10 (2015) 313-318.

[7] L. Yu, Q. Yan and A. Ruzsinszky, *Nat Commun* 8 (2017) 15224.

[8] P. Yu, X. Yu, W. Lu, H. Lin, L. Sun, K. Du, F. Liu, W. Fu, Q. Zeng and Z. Shen, *Adv. Fun. Mater* 26 (2016) 137-145.

[9] H. S. Song, S. L. Li, L. Gao, Y. Xu, K. Ueno, J. Tang, Y. B. Cheng and K. Tsukagoshi, *Nanoscale* 5(2013) 9666.

[10] G. Li, G. Ding and G. Gao, *J Phys : Condens Matter* 29 (2017) 015001.

[11] C. Ataca, H. Sahin and S. Ciraci, *J. Phys. Chem. C* 116 (2012) 8983.

[12] Z. Gao, F. Tao and J. Ren, *Nanoscale* 10 (2018) 12997-13003.

[13] Y. Wang, G. Qiu, R. Wang, S. Huang, Q. Wang, Y. Liu, Y. Du, W. A. Goddard, M. J. Kim, X. Xu, P. D. Ye and W. Wu, *Nat Electron* 1 (2018) 288.

[14] J. Chen, Y. Dai, Y. Ma, X. Dai, W. Ho and M. Xie, *Nanoscale* 9 (2017) 15945-15948.

[15] S. Keuleyan, M. Wang, F. R. Chung, J. Commons and K. J. Koski, *Nano Lett* 15 (2015) 2285-90.

[16] K. C. Göbgen, S. Steinberg and R. Dronskowski, *Inorg Chem* 56 (2017) 11398-11405.

[17] A. Kandemir, F. Iyikanat and H. Sahin, *J Phys : Condens Matter* 29 (2017) 395504.

[18] G. J. Snyder and E. S. Toberer, *Nat Mater* 7 (2008) 105-14.

[19] G. Zhang and Y. W. Zhang, *J. Mater. Chem. C* 5 (2017) 7684-7698.

[20] G. Kresse and J. Furthmüller, *Phys. Rev. B. Condens. Matter* 54 (1996) 11169-11186.

[21] G. Kresse and J. Furthmüller, *Comp. mat. er. sci* 6 (1996) 15-50.

[22] P. E. Blöchl, *Phys. Rev. B* 50 (1994) 17953-17979.

[23] G. Kresse and D. Joubert, *Phys. Rev. B* 59 (1999) 1758-1775.

[24] J. P. Perdew, K. Burke and M. Ernzerhof, *Phys. Rev. Lett* 77 (1998) 3865-3868.

[25] A. Togo, F. Oba and I. Tanaka, *Phys. Rev. B* 78 (2008) 134106.

[26] W. Li, L. Lindsay, D. A. Broido, D. A. Stewart and N. Mingo, *Phys. Rev. B* 86 (2012) 80-82.

[27] W. Li, J. Carrete, N. A. Katcho and N. Mingo, *Compu. Phys. Commun* 185 (2014) 1747-1758.

[28] G. K. H. Madsen and D. J. Singh, *Compu. Phys. Commun* 175 (2006) 67-71.

[29] Y. Ding, Y. Wang, J. Ni, L. Shi, S. Shi and W. Tang, *Phys. B. Phys. Condens. Matt* 406 (2011) 2254-2260.

[30] P. Rani and V. K. Jindal, *Rsc Adv* 3 (2012) 802-812.

[31] Q. Wei and X. Peng, *App. Phys. Lett* 104 (2014) 372-98.

[32] J. Zhou and R. Huang, *J. Mechan. Phys. Soli* 56 (2008) 1609-1623.

[33] Z. Gao, X. Dong, N. Li and J. Ren, *Nano Lett* 17 (2017) 772-777.

[34] X. Wu, V. Varshney, T. Lee, T. Zhang, J. L. Wohlwend, A. K. Roy and T. Luo, *Nano Lett* 16 (2016) 3925.

[35] Z. Gao, N. Li and B. Li, *Phys. Rev. E* 93 (2016) 022102.

[36] Z. Gao, N. Li and B. Li, *Phys. Rev. E* 93 (2016) 032130.

[37] X. J. Yan, Y. Y. Lv, L. Li, X. Li, S. H. Yao, Y. B. Chen, X. P. Liu, H. Lu, M. H. Lu and Y. F. Chen, *App. Phys. Lett* 110 (2017) 7805.

[38] Y. Hong et al., *Chinese Phys. B* 27 (2018) 036501.

[39] G. Zhang and Y. W. Zhang, *Chinese Phys. B 26* (2017) 034401.

[40] M. Dresselhaus, G. Chen, M. Tang, R. G. Yang, H. Lee, D. Z. Wang, Z. F. Ren, J. P. Fleurial and P. Gogna, *Adv. Mater* 19 (2007) 1043-1053.

[41] G. Zhang and Y. W. Zhang, *Mechanics of Materials* 91 (2015) 382-398.

[42] G. D. Mahan and J. O. Sofo, *PNAS* 93 (1996) 7436.

[43] Y. Pei, H. Wang and G. J. Snyder, *Adv Mater* 24 (2012) 6124-6124.

[44] J. Bardeen and W. Shockley, *Phys. Rev* 80 (1950) 69-71.

[45] X. Tan, H. Shao, T. Hu, G. Liu, J. Jiang and H. Jiang, *Phys. Chem. Chem. Phys* 17 (2015) 22872.

[46] H. Y. Lv, W. J. Lu, D. F. Shao, H. Y. Lu and Y. P. Sun, *J. Mater. Chem. C* 4 (2016) 4538-4545.

[47] L. Xu, M. Yang, S. J. Wang and Y. P. Feng, *Phys. Rev. B* 95 (2017) 235434.

[48] J. Xi, M. Long, L. Tang, D. Wang and Z. Shuai, *Nanoscale* 4 (2012) 4348.

[49] J. Qiao, X. Kong, Z. X. Hu, F. Yang and W. Ji, *Nat Commun* 5 (2014) 4475.

[50] G. Li, K. Yao and G. Y. Gao, *Nanotechnology* 29 (2018) 015204.

[51] L. C. Zhang, G. Qin, W. Z. Fang, H. J. Cui, Q. R. Zheng, Q. B. Yan and G. Su, *Sci. Rep* 6 (2018) 19830.

[52] Y. X. Zhen, M. Yang, H. Zhang, G. S. Fu, J. L. Wang, S. F. Wang and R. N. Wang, *Science. Bulletin* 22 (2017) 1530-1537.

[53] B. K. Ridley, *J. Phys. Chem. C. Solid State Phys* 15 (1982) 5899.

[54] X. Liu, J. Hu, C. Yue, N. D. Fera, Y. Ling, Z. Mao and J. Wei, *Acs Nano* 8 (2014) 10396-10402.

[55] O. Donmez, M. Gunes, A. Erol, C. M. Arikan, N. Balkan and W. J. Schaff, *Nanoscale. Research. Letters* 7 (2012) 490.

[56] M. S. Lee and S. D. Mahanti, *Phys. Rev. B Condens. Matter* 85 (2012) 1745-1751.

[57] Y. Takagiwa, Y. Pei, G. Pomrehn and G. J. Snyder, *APL. Mater* 1 (2013) 105.

[58] S. Sharma, N. Singh and U. Schwingenschlögl, *ACS Appl. Energy Mater* 1 (2018) 1950-1954.

[59] Z. Jie, X. L. Liu, Y. Wen, S. Lu, C. Rong, H. Liu and B. Shan, *ACS Appl. Mater and Inter* 9 (2017) 2509.

**List of figure captions**

**Fig. 1.** Structure diagram. (a) Top view of monolayer 1T-XTe$_2$ (X = Si, Sn). The dashed lines show the primitive cell. (b) and (c) are both the side view along the *x* and *y* direction respectively.

**Fig. 2.** Calculated phonon spectrum and DOS of monolayer 1T-type telluride structures. Phonon dispersions are shown along the high symmetry points Γ-M-K-Γ in the Brillouin Zone. (a)(b) monolayer SiTe$_2$ and (c)(d) monolayer SnTe$_2$.

**Fig. 3.** (a) Temperature dependence of lattice thermal conductivity for monolayer SiTe$_2$ and SnTe$_2$ from 300 K to 900 K. The dashed lines represent the fitting curves of 1/T. (b) Group velocity for monolayer SiTe$_2$ and SnTe$_2$. (c) Phonon lifetime of them at 300 K. (d) Phase space for three phonon scattering channels of them.

**Fig. 4.** Calculated electronic band structure and the total DOS under biaxial strain ε of monolayer SiTe$_2$ in figure (a) and (b), and figure (c) and (d) is monolayer SnTe$_2$ respectively. Bule, red and black lines represent ε= 5%, ε= -5% and no strain respectively. The dashed green lines represent the Fermi level. Electronic band structures of them are shown along high symmetry points Γ, M, K, Γ.

**Fig. 5.** The electronic transport parameters as a function of effective carrier concentration for moving Fermi level of monolayer SiTe$_2$ at 300 K, 600 K and 900 K. (a) Seebeck coefficient. (b) electronic conductivity. (c) the power factor *PF*. (d) the figure of merit *ZT*.

**Fig. 6.** The electronic transport parameters as a function of effective carrier concentration for moving Fermi level of monolayer SnTe$_2$ at 300 K, 600 K and 900 K. (a) Seebeck coefficient. (b) electronic conductivity. (c) the power factor *PF*. (d) the figure of merit *ZT*.

## Tables

**Table 1.** Optimized structure information of 1T-XTe$_2$ (X = Si, Sn). Lattice constant, the bond length of X-Te, the bond angle of X-Te-X. $d$ represents the thickness of structure. The intrinsic thickness of structures are shown outside the parentheses. Values in the parentheses are the vdW effective thickness of structures. $E_c$ is the cohesive energy in the unit cell.

|        | lattice constant (Å) | bond length (Å) | Bond angle (°) | Thickness $d$ (Å) | $E_c$ (eV) |
|--------|----------------------|------------------|----------------|---------------------|------------|
| SiTe$_2$ | 3.78 | 2.73 | 87.52 | 3.29 (7.77) | 15.24 |
| SnTe$_2$ | 4.10 | 2.99 | 86.82 | 3.63 (8.11) | 13.83 |

**Table 2.** The elastic modulus tensor $C_{ij}$ (GPa) are shown of monolayer SiTe$_2$ and SnTe$_2$. $E_x$ and $E_y$ are the in-plane Young's modulus, also called in-plane stiffness. $v_{xy}$ and $v_{yx}$ are the Poisson's ratio in the *x* and *y* directions.

|        | $C_{11}$ ($C_{22}$) | $C_{12}$ ($C_{21}$) | $C_{66}$ | $E_x$ (N/m) | $E_y$ (N/m) | $v_{xy}$ | $v_{yx}$ |
|--------|----------------------|----------------------|----------|---------------|---------------|----------|----------|
| SiTe$_2$ | 90.99 (90.99) | 63.68 (63.68) | 1.02 | 21.16 | 21.16 | 0.77 | 0.77 |
| SnTe$_2$ | 158.66 (158.66) | 122.33 (122.33) | 1.12 | 16.80 | 16.80 | 0.70 | 0.70 |

**Table 3.** The elastic modulus ($C_{2D}$), deformation potential constant ($E_l$), the effective mass ($m^*_{\Gamma-M}$, $m^*_{M-K}$), the mobility (μ) and relaxtion time $\tau_{ac}$ for acoustic phonon scattering, $\tau_{op}$ for polar optical phonon scattering and $\tau_{tot}$ is the total relaxtion time for monolayer SiTe$_2$ and SnTe$_2$ at 300 K. $m_e$ is the electron mass.

|        | $C_{2D}$ (J/m$^2$) | $E_l$ (eV) | $m^*_{\Gamma-M}$ ($m_e$) | $m^*_{M-K}$ ($m_e$) | $\mu_{ac}$ (cm$^2$/Vs) | $\tau_{ac}$ ($10^{-14}$s) | $\tau_{op}$ ($10^{-14}$s) | $\tau_{tot}$ ($10^{-14}$s) |
|--------|---------------------|-------------|----------------------------|----------------------|--------------------------|----------------------------|----------------------------|------------------------------|
| SiTe$_2$ | 52.20 | 6.98 | 1.56 | -0.16 | 692.17 | 2.60 | 6.26 | 1.84 |
| SnTe$_2$ | 32.98 | 4.73 | 0.46 | -0.21 | 535.63 | 5.79 | 2.14 | 1.56 |

**Figures**

Fig. 1

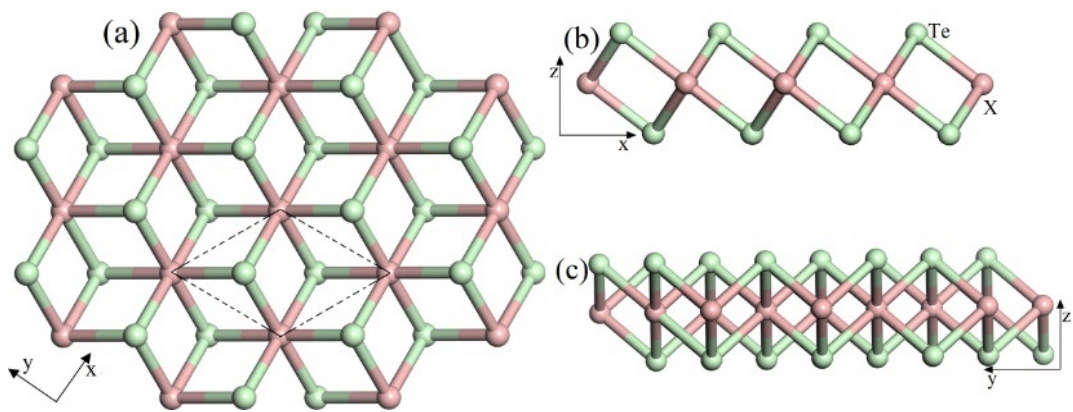

Fig. 2

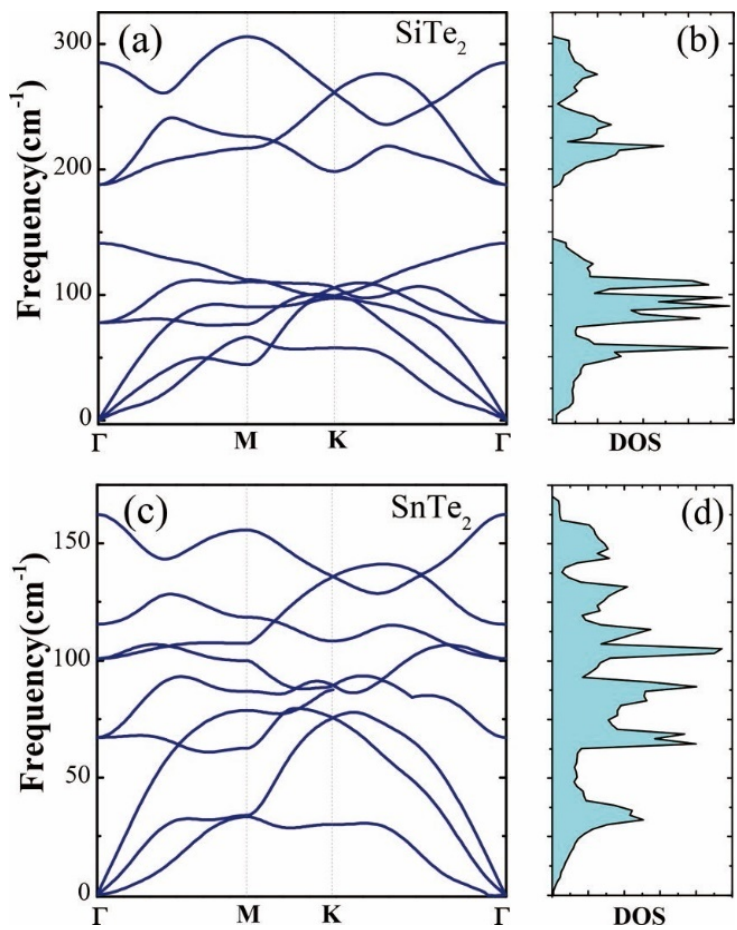

Fig. 3

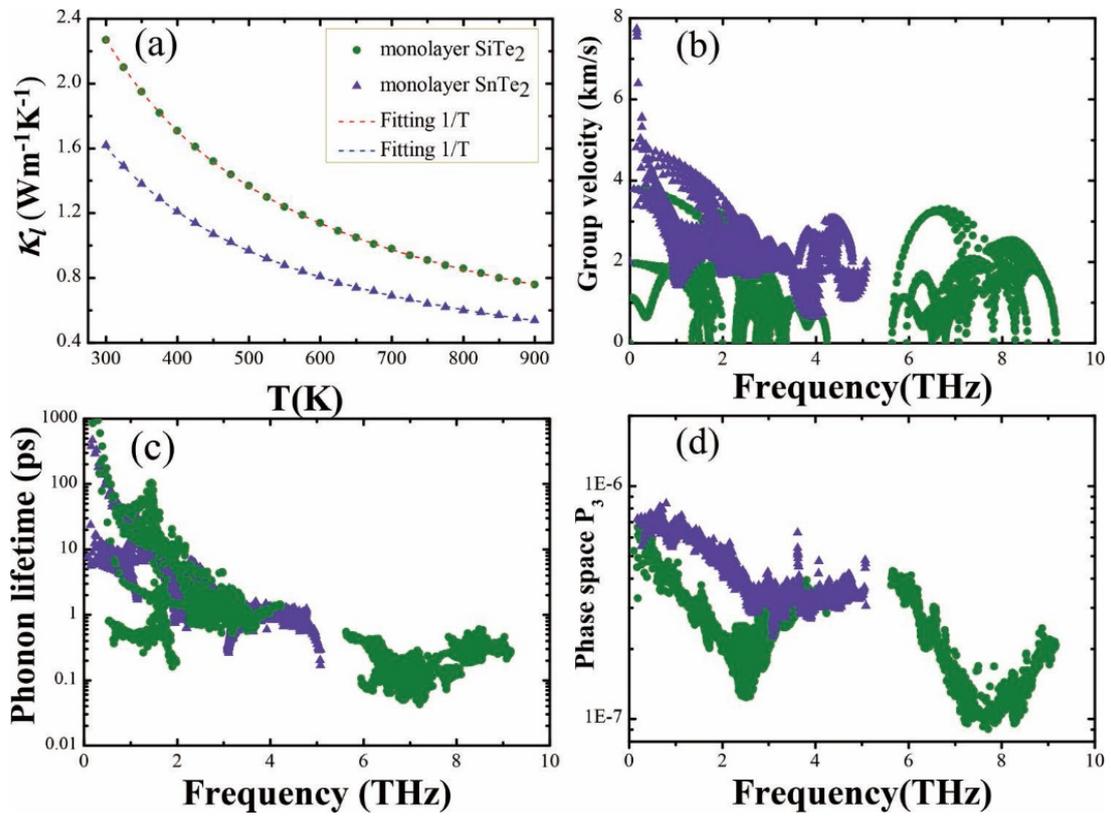

Fig. 4

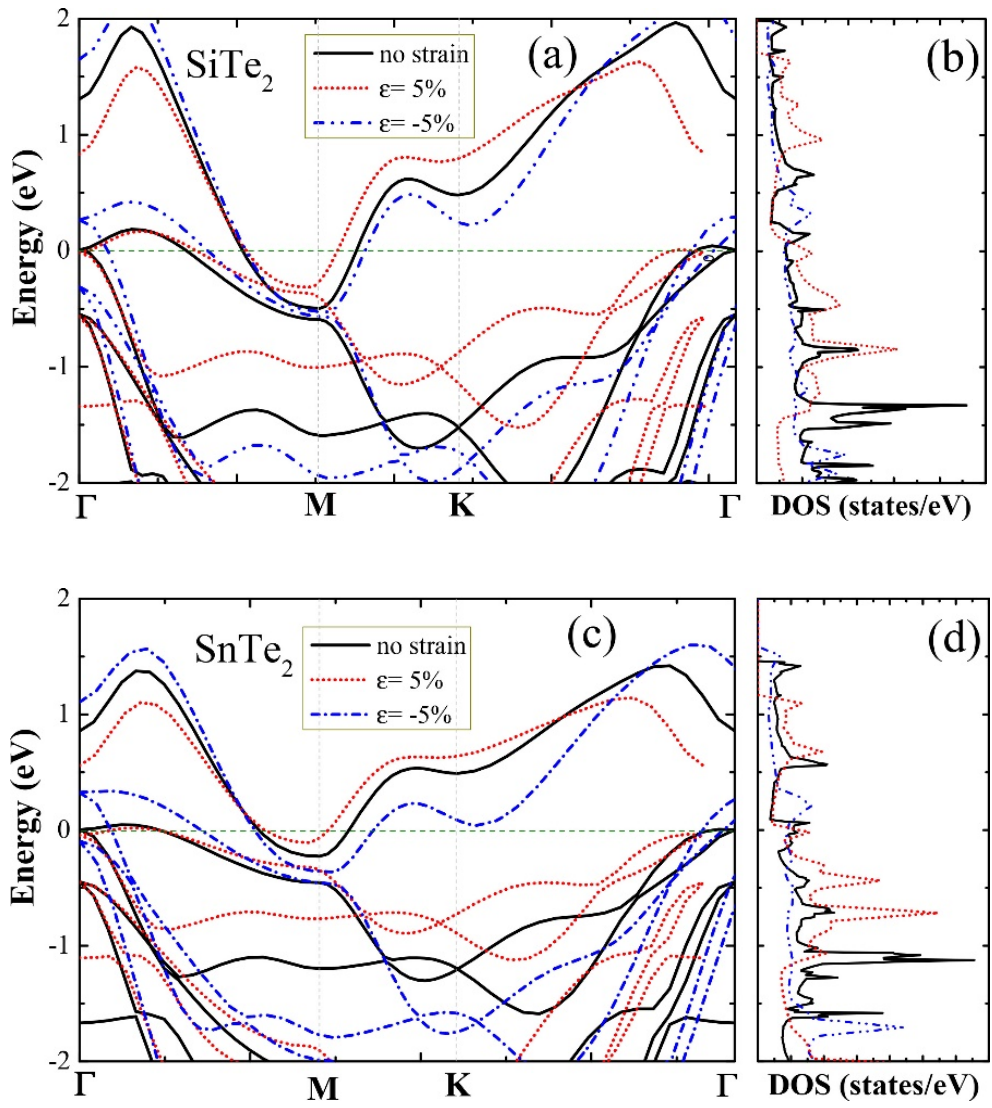

Fig. 5

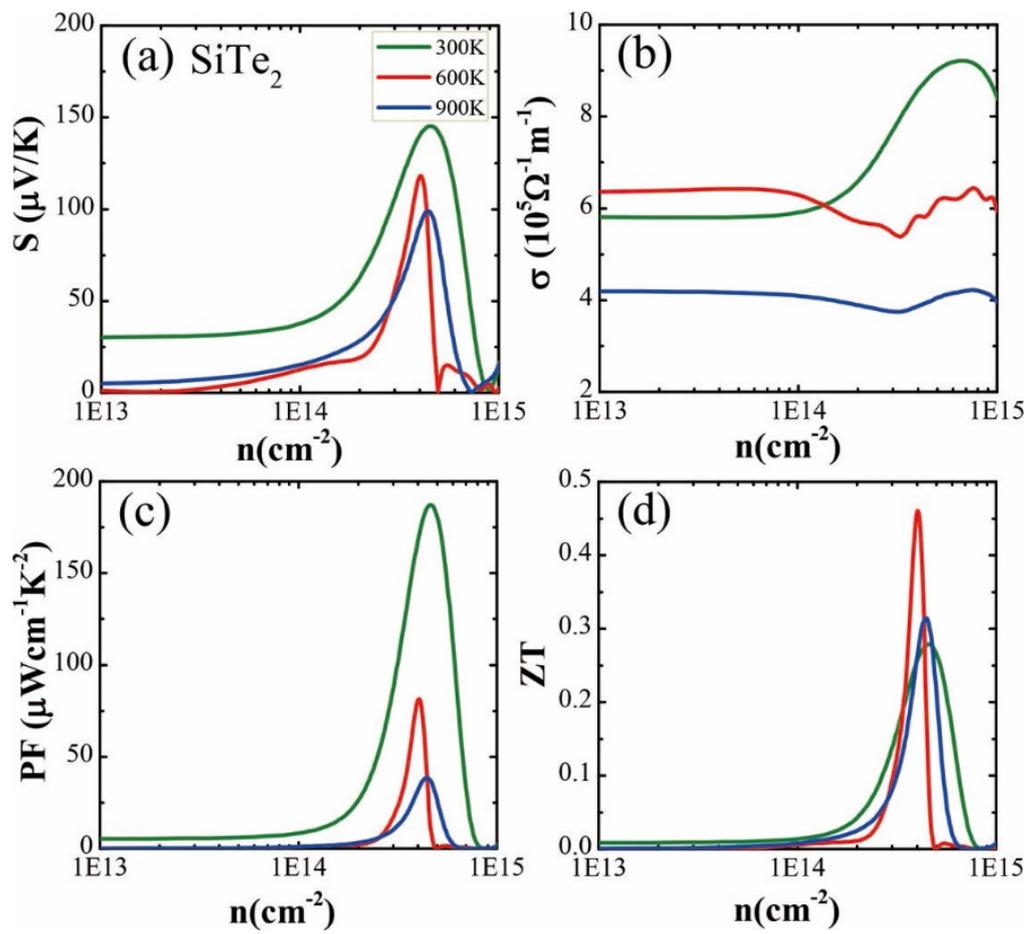

Fig. 6

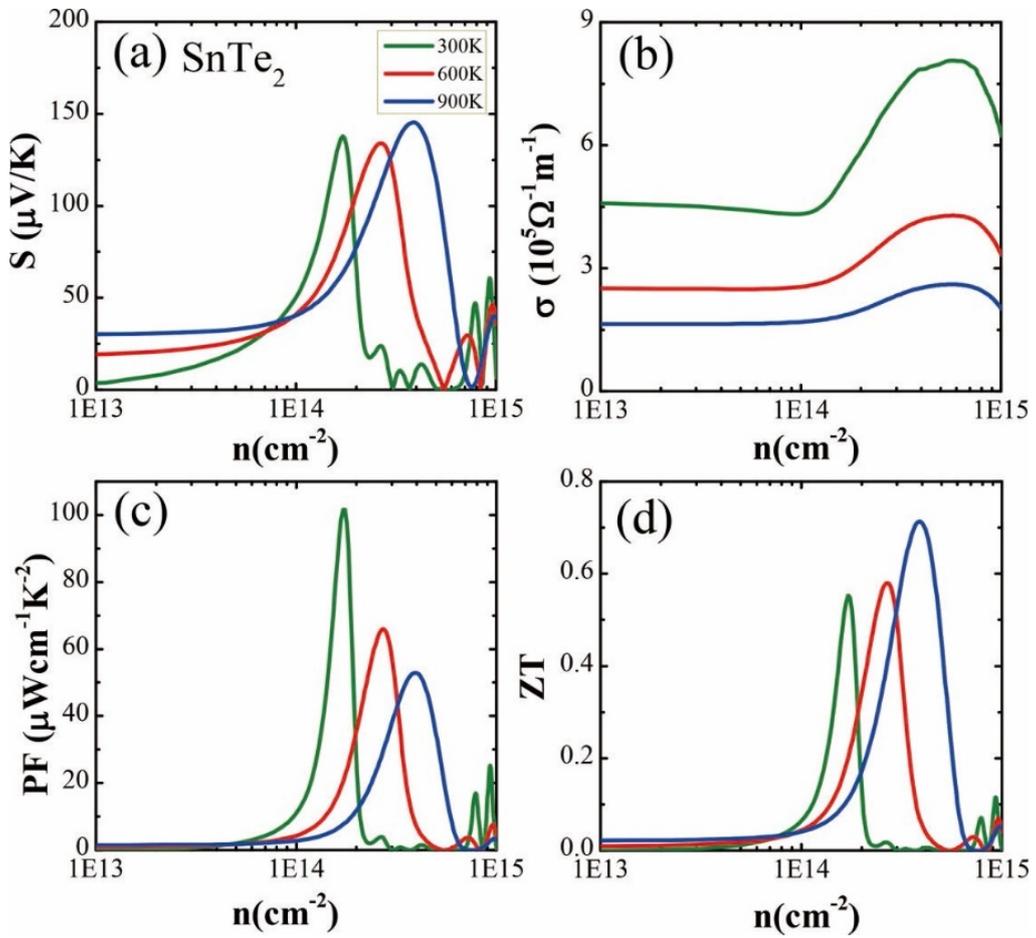

**Highlights:**

• Mechanical, electronic and thermoelectric properties of monolayer 1T phase semimetal $SiTe_2$ and $SnTe_2$ is firstly studied.

• Much smaller in-plane stiffness and ultralow thermal conductivity of monolayer $SiTe_2$ and $SnTe_2$ are obtained and physically explained.

• The two structures show relatively flexible electronic properties under large biaxial strain ε= ±5%, indicating potentially flexible electrode materials.

• We have clearly shown and discussed that both acoustic and polar optical phonon scattering play a critical role in the relaxation time of electronic properties in both 2D semimetal $SiTe_2$ and $SnTe_2$.